\documentclass[aps,prd,twocolumn,floats,showpacs, nofootinbib]{revtex4}
\usepackage{graphics,epsfig,color} 
\usepackage{amsmath,amssymb}
\usepackage{hyperref}

\def\d {\mbox{d}}
\def\L {\mathcal{L}}

\newcommand{\ul}[1]{\overline{#1}}

\newcommand{\w}[1]{\tilde{#1}}

\def\be{\begin{equation}}
\def\ee{\end{equation}}
\def\bea{\begin{eqnarray}}
\def\eea{\end{eqnarray}}
\begin{document}
\title{On the cosmic convergence mechanism of the massless dilaton}
\author{Olivier Minazzoli}
\affiliation{UMR ARTEMIS, CNRS, University of Nice Sophia-Antipolis,
Observatoire de la C\^ote d’Azur, BP4229, 06304, Nice Cedex 4, France}

\begin{abstract}
The converging mechanism discussed in [Damour \& Nordtvedt, Physical Review Letters,70,15] for scalar-tensor theories has been applied to dilaton-like theories in several subsequent papers. In the present communication, we show that an unfortunate assumption in those studies led to a scalar-field equation unsuitable for the study of the dilaton field. The corrected scalar-field equation turns to change the numerical outcome of those studies in general, but even sometimes their qualitative aftermath. Therefore, the present result call for new investigations of the subject. On the other hand, our result shows that the string-inspired theory presented in [Minazzoli \& Hees, Physical Review D,88,4] is naturally solution to the problem of the effective constancy of the fundamental coupling constants at late cosmic times, while it requires less fine-tuning than other massless dilaton or usual stalar-tensor theories.

\end{abstract}
\maketitle

{\it Introduction}.--- 
In perturbative string theory, the effective action of gravitation is not general relativity but a scalar-tensor theory with non-minimal coupling between the scalar-field (called \textit{dilaton}) and both the Ricci scalar and the material part of the Lagrangian \cite{callanNuPhB86,Sstring88,eastherPRD96,gasperiniPRD02,DamPolyGRG94,DamPolyNPB94,damourPRL02,damourPRD02,damourPRD10}. While the former coupling is roughly speaking mainly constrained by solar system observations of the gravitational post-Newtonian phenomenology, the latter is much more severely restricted by the present tight constraints on equivalence principle violations. Hence, in order to reconcile perturbative string theory with the present strong constraints on the equivalence principle, it has been postulated that the dilaton field would acquire a mass term through non-perturbative effects (see for instance \cite{gasperiniPRD02} and references therein). Indeed, such a mass term would freeze the dilaton dynamic at macroscopic scales, hence leading to an effective satisfaction of the various properties following from the equivalence principle (eg. constancy of the coupling constants). However, a decoupling mechanism has also been found \cite{DamPolyGRG94,DamPolyNPB94}, that does not need any dilaton self-interacting potential. The mechanism turned out to be almost the same as in usual scalar-tensor theories \cite{damourPRL93,damourPRD93}.

In the present communication we demonstrate that the field equations used by \cite{DamPolyGRG94,DamPolyNPB94} are incomplete. The omission can be explained by an unfortunate assumption they used, that we already noticed in \cite{moiPRD13_2} while studying the massless dilaton post-Newtonian phenomenology. Talking about their assumption, Damour and Polyakov say ``We believe that our main qualitative conclusions do not depend strongly on the specific form of the assumption (2.11)''. In the following, we show that their assumption actually fails to predict the correct scalar-field equations. It has to be noted that not considering Damour and Polyakov's assumption also allowed the authors of \cite{nousPRD13} to find a string-inspired theory that passes solar-system tests with flying colors, even for a massless scalar-field. The theory phenomenologically proposed in \cite{nousPRD13} is therefore an alternative to the usual non-perturbative mass assumption.

{\it Derivation of the result}.--- 
In order to simplify the discussion, we shall consider that the dilaton couples universally to the matter Lagrangian\footnote{Such a condition seems to be necessary for the driving mechanism to occur \cite{DamPolyGRG94,DamPolyNPB94}.}. This assumption does not impact the outcome of the study. Hence, let us start with the action of a class of scalar-tensor theory with universal coupling between the scalar field and the material Lagrangian, directly given in 4 dimensions such as in \cite{DamPolyGRG94,DamPolyNPB94,damourPRD96,gasperiniPRD02,damourPRL02,damourPRD02,damourPRD10}:
\begin{eqnarray}
S= \frac{1}{c}\int&&  d^4x \sqrt{-g} ~\frac{1}{2\mu}\times \label{eq:actiondila}  \\ 
&&\left[\Phi  R-\frac{\omega(\Phi)}{\Phi} (\partial_\sigma \Phi)^2+ 2 \mu~f(\Phi)~\mathcal{L}_m (g_{\mu \nu}, \Psi) \right] . \nonumber
\end{eqnarray}
where $g$ is the metric determinant, $R$ is the Ricci scalar constructed from the metric $g_{\alpha \beta}$, $\mu$ is a coupling constant with the dimensions $L^{-1} M^{-1}T^2$, $f(\Phi)$ is an arbitrary non-dimensional differentiable and real function of $\Phi$, $\mathcal{L}_m$ is the material Lagrangian and $\Psi$ represents the non-gravitational fields. It has to be noted that such an action encompasses the effective string theory low energy action considered in \cite{DamPolyGRG94,DamPolyNPB94} (for which $f(\Phi) \propto \Phi$), as well as the string-inspired theory proposed in \cite{nousPRD13} (for which $f(\Phi) \propto \sqrt{\Phi}$) \footnote{Though unlikely, it has to be noted that such a coupling function could result from the non-perturbative effective action of string theory. The corresponding scalar-field has been dubbed ``\textit{pressuron}'' in \cite{moiARXIV2014}, because it decouples in pressure-less regimes.}. Indeed, although the form is slightly different from the one given for instance in \cite{DamPolyGRG94,DamPolyNPB94}, the differences reduce to a rescaling of the scalar-field and to a total derivative term that does not contribute to the field equations. We prefer our present notation because it is similar to the one found usually in scalar-tensor literature.

Following \cite{DamPolyGRG94,DamPolyNPB94}, the action can be reformulated in the so-called Einstein representation\footnote{Also know as Einstein \textit{frame}.}. The action can be written as
\begin{eqnarray}
S=\frac{1}{2} \int d^4x\sqrt{-\tilde{g}} \left(  \tilde{R} - \tilde{g}^{\alpha \beta} \partial_\alpha \ul{\varphi} \partial_\beta \ul{\varphi} \right) + S_m, \label{eq:SEwRS}
\end{eqnarray}
where we have set $\mu =c=1$ for simplicity, with
\begin{eqnarray}
S_m&=& \frac{1}{2}\int d^4x\sqrt{-g}~ 2 f(\Phi) \mathcal{L}_m(g_{\mu \nu}, \Psi),\label{eq:actionMJ}\\
&=& \frac{1}{2} \int d^4x\sqrt{-\tilde{g}} ~2 f(\Phi(\ul{\varphi})) \tilde{\mathcal{L}}_m(\w{g}_{\nu \nu}, \Phi(\ul{\varphi}), \Psi).
\end{eqnarray}
where $g^{\alpha \beta} \equiv \Phi \tilde{g}^{\alpha \beta}$, $\tilde{\mathcal{L}}_m=\Phi^{-2}\mathcal{L}_m$, $\sqrt{-g}= \Phi^{-2} \sqrt{-\tilde{g}} $, $\varphi \equiv \ln \Phi$ and $\d \ul{\varphi}=\pm \sqrt{\omega + 3/2} ~\d \varphi$.\footnote{It is important to note that since the redefinition of the scalar field is not differentiable at the general relativity limit (ie. $\omega \rightarrow \infty$), the equivalence between the Einstein representation action defined by equation (\ref{eq:SEwRS}) and the string representation action (\ref{eq:actiondila}) is lost at this limit \cite{jarvPRD07}. Therefore, a better choice would be to work with a non-rescaled scalar field such as the one defined by the action (B.2) in \cite{moiPRD13_2} or (5) in \cite{jarvPRD07}.} The variation of the material sector through relevant fields therefore writes\footnote{See appendix B in \cite{moiPRD13_2} for a derivation of this result.}
\begin{eqnarray}
\delta S_m=\frac{1}{2}\int d^4x\sqrt{-\tilde{g}} \big(&-& f \tilde{T}_{\alpha \beta}~ \delta \tilde{g}^{\alpha \beta} \label{eq:deltaSM2} \\
&+&2 \left[\alpha f \w{T} + f_{, \ul{\varphi}} \w{\L}_m \right] \delta \ul{\varphi} \big),\nonumber
\end{eqnarray}
where
\be
\alpha(\ul{\varphi}) \equiv -\frac{1}{2} \frac{\partial \ln \Phi}{\partial \ul{\varphi}} = -\frac{1}{2} \frac{\partial \varphi}{\partial \ul{\varphi}}=\mp \frac{1}{2\sqrt{\omega(\ul{\varphi})+3/2}}, \label{eq:alpha}
\ee
and
\be
T_{\alpha \beta} \equiv - \frac{2}{\sqrt{-g}} \frac{\delta \left(\sqrt{-g} \L_m \right)}{\delta g^{\alpha \beta}},
\ee
with $\omega(\ul{\varphi}) \equiv \omega(\Phi(\ul{\varphi}))$, $\tilde{T}_{\alpha \beta}=\Phi^{-1}T_{\alpha \beta}$ and $\w{T}=\Phi^{-2} T$ ($\tilde{T}^{\alpha \beta}=\Phi^{-3}T^{\alpha \beta}$). Equation (\ref{eq:deltaSM2}) is particularly important because it gives the source of the field equations. In particular the source $\ul{\sigma}$ of the scalar-field, such that $\ul{\sigma}= (-\tilde{g})^{-1/2} \delta S_m/\delta \ul{\varphi}$, is given by the second term in the right hand side of equation (\ref{eq:deltaSM2}). Instead of deriving $\ul{\sigma}$ from the string representation to the Einstein representation as we just did, \cite{DamPolyGRG94,DamPolyNPB94} directly work with a non-interactive point particle action written in the Einstein representation
\be
S^{DamPoly}_m= - \sum_A \int \w{m}_A(\ul{\varphi}) d\w{s},
\ee
where $d\w{s}^2=\w{g}_{\alpha \beta} dx^\alpha dx^\beta$. Then, they  assume a dilaton functional dependency of the masses in the Einstein representation $\w{m}_A(\ul{\varphi})$ \cite{DamPolyGRG94,DamPolyNPB94} and subsequently deduce the sought-after $\ul{\sigma}$. By doing so, as explained in \cite{moiPRD13_2}, Damour and Polyakov \cite{DamPolyGRG94,DamPolyNPB94} miss the fact that the scalar-field coupling is no longer simply related to the conformal factor $\Phi$ through the function $\alpha$ (\ref{eq:alpha}), but also depends on the gradient of the coupling function $(\ln f)_{,\ul{\varphi}}$, as well as the material Lagrangian $\w{\L}_m$, in a non-trivial way (\ref{eq:deltaSM2}). 
The equations resulting from the action in the Einstein representation (\ref{eq:SEwRS}) are
\bea
\w{R}_{\alpha \beta}-\frac{1}{2} \w{g}_{\alpha \beta} \w{R} &=& f \w{T}_{\alpha \beta} \\
&+& \partial_\alpha \ul{\varphi} \partial_\beta \ul{\varphi}- \frac{1}{2} \w{g}_{\alpha \beta} \w{g}^{\sigma \epsilon} \partial_\sigma \ul{\varphi} \partial_\epsilon \ul{\varphi}, \nonumber
\eea
and
\be
\w{\Box} \ul{\varphi} = - \alpha~f \w{T} -  f_{, \ul{\varphi}} ~\w{\L}_m,\label{eq:dilatonvr}
\ee
where the tilde on the operator refers to the fact that it is constructed with the metric $\w{g}_{\alpha \beta}$. One can notice that the last term of equation (\ref{eq:dilatonvr}) is missing in \cite{DamPolyGRG94,DamPolyNPB94,damourPRL02,damourPRD02,damourPRD10,damourCQG12}. This oversight can be directly imputed to the assumption they used on the functional dependency of the masses in the Einstein representation. At the same time, the invariance of the action (\ref{eq:SEwRS}) under diffeomorphism induces the following conservation equation
\be
\w{\nabla}_\sigma \w{T}^{\alpha \sigma}= \alpha \w{T} ~\w{\nabla}^\alpha   \ul{\varphi} + \left( \w{g}^{\alpha \sigma} \w{\L}_m -  \w{T}^{\alpha \sigma} \right) \partial_\sigma \ln f. \label{eq:conservE}
\ee
It has to be noted this equation differs from the usual scalar-tensor case for $f \neq$ Cste. Moreover, it has to be pointed out that the last term of the right hand side in equation (\ref{eq:conservE}) is missing in \cite{DamPolyGRG94,DamPolyNPB94} as well. However, in the specific case of a perfect fluid in a matter dominated Friedmann universe, it turns out that this term vanishes due to an exact cancellation \cite{moiARXIV2014}. Therefore in matter dominated Friedmann universes, the conservation equation is remarkably the same as in usual scalar-tensor theories. On the contrary, in post-Newtonian developments, such a term is non-null and plays an important role (see eg. \cite{nousPRD13}).  

As in \cite{DamPolyGRG94,DamPolyNPB94}, let us now consider the perfect fluid approximation, such that $\w{\L}_m= - \w{\epsilon}$ \cite{moiPRD12,moiPRD13} and $\w{T}=- \w{\epsilon}+3\w{P}$, where $\w{\epsilon}$ and $\w{P}$ are respectively the total energy density and the pressure of the fluid in the Einstein representation. Let us note, however, that there is no reason to assume that the effective macroscopic perfect fluid Lagrangian $\w{\L}_m= - \w{\epsilon}$  is also valid for the various imperfect fluids that drive the radiation period. Otherwise, let us restrict our attention to the simple case $f(\Phi) \propto \Phi^n$, with $n \in \mathbb{R}$, such that $n=1$ corresponds to the theory treated in \cite{DamPolyGRG94}, and $n=1/2$ to the string-inspired theory treated by \cite{nousPRD13}. The scalar-field equation then reduces to
\be
\w{\Box} \ul{\varphi} = \alpha f \left[(1-2n)  \w{\epsilon} - 3 \w{P} \right].
\ee
Assuming a flat Friedmann universe, one can find a specific evolution parameter $p$ such that the scalar-field equation is independent of the cosmic scale factor \cite{damourPRL93,damourPRD93}\footnote{One should note that this property is not an exclusive feature of the Einstein representation \cite{sernaCQG02}.}. Indeed, defining $p = \ln a + \rm{Cste}$, one gets the decoupled scalar-field equation
\be
\frac{ \ul{\varphi}''}{3- \ul{\varphi}'^2/2} +\frac{1}{2}\left(1-\frac{\w{P}}{\w{\epsilon}} \right)  \ul{\varphi}'= - \left(1-2n-3 \frac{\w{P}}{\w{\epsilon}} \right) \alpha(\ul{\varphi}), \label{eq:STFedecoupled}
\ee
where $X' \equiv dX/dp$ \footnote{It is remarkable that the function $f$ appears through the ratio $\Phi f_{,\Phi}/f=n$ only. Otherwise, for $n=0$, our equation differs slightly from (3.15) in \cite{damourPRD93} because of the sightly different choice of scalar-field rescaling. Indeed, conversely to them, we chose not to have a factor 2 in front of the scalar-field kinetic term in equation (\ref{eq:SEwRS}). But the two results are of course equivalent in the limit $n=0$.}. 

From the last term of (\ref{eq:dilatonvr}), one can see that the dilaton has a source even when $T=0$, for $\w{\L}_m \neq 0$. However, the whole converging mechanism described in \cite{DamPolyGRG94,DamPolyNPB94} --- and then subsequently used in \cite{damourPRL93,damourPRD93,damourPRL02,damourPRD02,damourPRD10,damourCQG12} --- lies on the no-source property of the scalar-field during the radiation era, because it allows to damp away any preradiation-era dynamic. Without this property, one can expect a stronger dependence on initial conditions and on the actual shape of $\alpha$. Therefore one should expect a different behavior of the scalar-field. Even if there is still convergence, one should not expect that the so-called \textit{attracting power} of the radiation era ($F_r$ in \cite{DamPolyNPB94}) has the same magnitude as in a no-source case. 
Now regarding the matter era, let us note that the sign of the source term changes in (\ref{eq:STFedecoupled}), whether one considers theories with $n>1/2$ or $n \in ~] 0; 1/2 [$. Such a sign can be directly related to the sign of the post-Newtonian parameter $\gamma$ \cite{moiPRD13_2}. Therefore, solutions will be qualitatively different whether one considers theories with $n>1/2$ or $n \in ~] 0; 1/2 [$, and more generally depending on $(\ln f)_{,\ul{\varphi}}$. In particular, for $n \in ~] 0; 1/2 [$ --- implying $\gamma<1$ \footnote{while $n>1/2$ and $n=1/2$, imply $\gamma >1$ and $\gamma=1$ respectively \cite{moiPRD13_2}.} --- if the dilaton converges towards $|\alpha|_{min}$ for $\epsilon \sim 3 P$, it diverges for $P \sim 0$, because the attracting force (ie. rhs. of equation (\ref{eq:STFedecoupled})) becomes repulsive (and the other way around). However, one has to keep in mind that one cannot simply use the approximation $\epsilon \sim 3 P$ in order to describe the radiation period because one doesn't know the effective macroscopic Lagrangian $\w{\L}_m$ of the imperfect fluids that drive this era.

{\it Conclusion}.--- 
We do not say that the converging mechanism described by Damour and Polyakov can no longer occur, but we argue that it depends more firmly than previously thought on initial conditions and on the specificity of the theory considered (through the functions $\alpha$ and $n=\Phi (\ln f)_{,\Phi}$). In any case, the whole problem of the massless dilaton cosmology, its convergence toward general relativity, and the magnitude of the expected equivalence principle violations should be worked anew with the correct scalar-field equations given by (\ref{eq:dilatonvr}-\ref{eq:conservE}).

Otherwise, one should notice that the exponential damping that may no-longer occur during the radiation era, occurs during the matter era for all $\alpha$ (thus $\omega$) if $n=1/2$ --- corresponding to the theory recently proposed in \cite{nousPRD13}. Therefore, the theory presented in \cite{nousPRD13} is naturally solution to the problem of the effective constancy of the fundamental coupling constants at late cosmic times, while it also passes solar system post-Newtonian tests with flying colors for all $\omega$ not too close of the singular value $-3/2$ \cite{nousPRD13} (resp. $\alpha \rightarrow \pm \infty$ (\ref{eq:alpha})). Hence, conversely to other massless dilaton theories and to usual scalar-tensor theories, the theory treated in \cite{nousPRD13} may not need the fine-tuned requirement that the ``late times'' local minimum of the function $\alpha$ is zero. Indeed, in order to satisfy current solar-system tests, theories with $n=1/2$ don't need to converge towards $\alpha_{min}=0$ because they decouple in the matter era anyway. Nevertheless, they are still able to converge towards $|\alpha|_{min}$ --- whether it is zero or not --- during the radiation era. The decoupling then comes by mass threshold: each time the universe passes a threshold $kT_i \sim m_i c^2$ when it cools down, the quantity $P/\epsilon$ declines up to a value of order $(m_i/kT)^2$, where $m_i$ is the mass of the species of particle/antiparticle `i' \cite{damourPRL93,damourPRD93}. In some sense, theories with $n=1/2$ have a reversed cosmology compared to usual scalar-tensor theories studied in \cite{damourPRL93,damourPRD93}. Testing the theory in strong regimes should give more constraints on the value of $\alpha$ at present time by constraining $\omega$ more tightly than it is with solar-system tests \cite{nousPRD13}.

The details of the scalar-field cosmic evolution shall be presented in dedicated communications.

\pagebreak

\bibliography{RNC}

\begin{acknowledgments}
The author wants to thank Aur\'elien Hees, Salvatore Capozziello and Viktor Toth for their interesting comments.
\end{acknowledgments}

\end{document}